\documentclass{emulateapj}

\usepackage{xcolor}
\usepackage{graphicx}
\usepackage{lineno}

\newcommand{\dpar}[2][\;\;]{\ensuremath{ \frac{\partial{#1}}{\partial{#2}} }}
\newcommand{\dparn}[3][\;\;]{\ensuremath{ \frac{\partial^{#3}{#1}}{\partial{#2}^{#3}} }}

\newcommand{\bvec}[1]{{\mbox{{\boldmath$#1$}}}}		
\newcommand{\unitv}[1]{\bvec{\hat{#1}}}			
\newcommand{\grad}{\bvec{\nabla}}			

\newcommand{\lrsp}{LRSP}	
\newcommand{\ptrslA}{RSPTA}	
\newcommand{\sci}{Sci}		

\newcommand{\D}{\displaystyle}
\newcommand{\eqnref}[1]{(\ref{#1})}

\begin{document}

\title{Do Coronal Loops Oscillate in Isolation?}
\shorttitle{Do Coronal Loops Oscillate in Isolation?}

\author{Bradley W. Hindman}
\affil{JILA, University of Colorado, Boulder, CO~80309-0440, USA}
\affil{Department of Applied Mathematics, University of Colorado, Boulder, CO~80309-0526, USA}
\email{hindman@solarz.colorado.edu}

\author{Rekha Jain}
\affil{School of Mathematics \& Statistics, University of Sheffield, Sheffield S3 7RH, UK}

\accepted{for publication by the Astrophysical Journal}


\begin{abstract}

Images of the solar corona by extreme-ultraviolet (EUV) telescopes reveal elegant
arches of glowing plasma that trace the corona's magnetic field. Typically, these
loops are preferentially illuminated segments of an arcade of vaulted field lines
and such loops are often observed to sway in response to nearby solar flares. A
flurry of observational and theoretical effort has been devoted to the exploitation
of these oscillations with the grand hope that seismic techniques might be used
as probes of the strength and structure of the corona's magnetic field. The commonly
accepted viewpoint is that each visible loop oscillates as an independent entity
and acts as a one-dimensional (1D) wave cavity for magnetohydrodynamic (MHD) kink
waves. We argue that for many events, this generally accepted model for the wave
cavity is fundamentally flawed. In particular, the 3D magnetic arcade in which the
bright loop resides participates in the oscillation. Thus, the true wave cavity is
larger than the individual loop and inherently multidimensional. We derive the skin
depth of the near-field response for an oscillating loop and demonstrate that most
loops are too close to other magnetic structures to oscillate in isolation. Further,
we present a simple model of a loop embedded within an arcade and explore how the
eigenmodes of the arcade and the eigenmodes of the loop become coupled. In particular,
we discuss how distinguishing between these two types of modes can be difficult when
the motions within the arcade are often invisible.

\end{abstract}



\section{Introduction}
\label{sec:Introduction}

The bright coronal loops that manifest in EUV observations are often observed to
vacillate in the aftermath of a solar flare \citep[i.e.,][]{Aschwanden:1999,
Nakariakov:1999, Wills-Davey:1999}. More recently it has been discovered that many
loops constantly tremble at low-amplitude, even in the absence of large excitation
events such as flares and coronal mass ejections \citep{Nistico:2013, Anfinogentov:2013,
Wang:2012, Duckenfield:2018}. The suggestion that the observed loop oscillations
are resonant  magnetohydrodynamic (MHD) waves and that the frequencies of these
resonant modes might be used as a seismic probe of the corona \citep{Roberts:1984}
has spurred the development of the entire discipline of coronal seismology---see
the review by \cite{Nakariakov:2020}.
 
The seminal works of \cite{Edwin:1983} and \cite{Roberts:1984} have become the
cornerstone of coronal seismology. These studies theoretically examine wave propagation
in coronal loops and highlight the important connection of these waves to coronal
observations. The models introduced in these papers popularized the approximation
of a coronal loop as a slender, straight, magnetic tube and, in so doing, have become
the theoretical backbone of seismic analysis of the properties of coronal loops.
Under such a thin-tube assumption, there are two forms of wave that predominate:
the kink (transverse) and sausage (longitudinal) oscillations. It is believed that
the observed vacillation of coronal loops is due to standing kink waves trapped in
an essentially 1D cavity lying between the two foot points of the loop in the photosphere
\citep[i.e.,][]{Aschwanden:1999, Nakariakov:1999, Goddard:2016}

The model of \cite{Edwin:1983} demonstrates that when the magnetic flux tube is
thin, the kink wave propates along the loop at a well-defined phase speed, $c_k$,
called the kink speed,

\begin{equation} \label{eqn:kink_speed}
	c_k^2 = \frac{B_0^2 + B_e^2}{4\pi\left(\rho_0 + \rho_e\right)} \; .
\end{equation}

\noindent In this expression, $\rho_0$ and $\rho_e$ are, respectively, the mass
densities inside the loop and in the coronal environment surrounding the loop.
Similarly, $B_0$ and $B_e$ are the magnetic field strength inside and outside the
loop. Since undulating coronal loops are thought to be overdense with $\rho_0 \gg \rho_e$
\citep{Lenz:1999, Winebarger:2003, Reale:2014} and $B_0\approx B_e$, the Alfv\'en
speed in the loop is less than the Alfv\'en speed in the surrounding corona, and
the kink speed is intermediate, with a value between the two Alfv\'en speeds. The
goal of coronal seismology is to measure the frequency $\omega$ and wavenumber $k_z$
of the oscillation and thus deduce the kink speed, $c_k = \omega/k_z$, which provides
a weighted average of the magnetic field strength.

\newpage

The model of \cite{Edwin:1983} makes four fundamental assumptions:
\begin{itemize}
	\item The curvature of the coronal loop's axis can be ignored.
	\item The loop and the surrounding corona are invariant along the loop.  
	\item The coronal loop lacks substructure in radius and azimuth that
		modify the kink wave frequency.
	\item The corona surrounding the loop lacks structure that causes 
		reflections.
\end{itemize}

\begin{figure*}
	\epsscale{0.5}
	\plotone{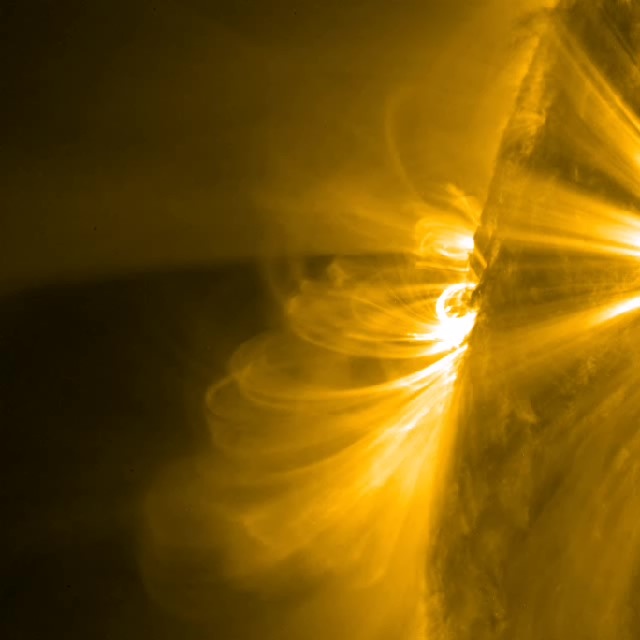}

	\caption{\small A still image that presents the first frame of a movie sequence
that is provided as a separate ancillary file (figure1.mp4). The still and the animation
show an arcade of magnetic field lines in the corona located on the solar limb. The sequence
of images was obtained by AIA in the Fe IX 171 \AA~ bandpass. Just after the animation
begins, a flare occurs causing the arcade to spasm.  The animation clearly demonstrates
that the entire arcade and the larger active region participate in the oscillation and
that MHD fast waves propagate across magnetic field lines.
	\label{fig:movie}}
\end{figure*}

\noindent The implications of the first three of these assumptions have been
studied extensively. The frequency shifts and polarization caused by loop
curvature have been explored and characterized \citep[e.g.,][]{Terradas:2006,
Gruszecki:2007, Ruderman:2009, VanDoorsselaere:2009, Pascoe:2014, Hindman:2015,
Thackray:2017}. Frequency shifts to the various longitudinal overtones have
been exploited as a means to assess variation in the properties of the loop
along its length \citep{Andries:2005, Goossens:2006, Dymova:2006, Diaz:2007,
McEwan:2008, Ruderman:2008, Verth:2008, Orza:2012, Jain:2012}. Theoretical
work has shown that closely packed magnetic threads oscillate as a collective
entity and the resulting kink speed is an average of the properties of all of
the threads \citep[e.g.,][]{Terradas:2008}. Conversely, the importance and
applicability of the fourth assumption have not received sufficient attention.
Observations suggest that the oscillation of a loop is often coupled to the
oscillation of nearby magnetic structures \citep{Schrijver:2000, Schrijver:2002,
Verwichte:2004, Verwichte:2009, Jain:2015, Li:2017}. The animation appearing
as Figure~\ref{fig:movie} illustrates the problem. This sequence of images
was obtained by the Atmospheric Imaging Assembly (AIA) on board the
{\it Solar Dynamics Observatory} (SDO) using the Fe IX 171 \AA~ bandpass.
After a flare, MHD waves can be seen to propagate back and forth across the
entire active region. Since, the waves propagate across field lines they must
be fast magnetic-pressure waves instead of pure kink waves. Further, the coronal
loops are not the only features that sways back and forth. There are correlated
motions of the loops, of the arcade of field lines in which the loops are embedded,
and in fact of the entire active region. From movies like Figure~\ref{fig:movie}
it is clear that coronal loops rarely oscillate in isolation.

Here, we will coarsely examine some of the issues associated with structuring of
the environment around the loop. In particular we hope to address the question,
``Under what conditions can a coronal loop oscillate in isolation, such that the
standard model of loop oscillations can be applied?" In order to assess these
issues, we need to consider two possibilities that can occur simultaneously for
waves of different frequency: evanescence or propagation within the coronal environment
and the host arcade. We will examine these two possibilities separately in sections
2 and 3. In section 4, we will present our conclusions and discuss the
implications of those conclusions on the reliability of coronal seismology.


\section{Evanescent Waves in the External Fluid: Skin Depth}
\label{sec:SkinDepth}

In order for kink waves to be a resonant oscillation of a coronal loop, the MHD
wave's energy density must be strongly concentrated within the loop or in the
immediate environs of the loop. Usually, this is accomplished by the wave being
evanescent outside the loop. But even if evanescent, the wavefield penetrates some
distance into the fluid surrounding the loop in a region called the near field.
The exponential decay length of this penetration, or the {\sl skin depth}, can be
derived without a detailed consideration of the loop's geometry and cross section.
The result depends on the Alfv\'en speed of the external media and the temporal
frequency of the wave. We will perform this calculation here and then focus on
thin flux tubes with a circular cross section.

\subsection{Derivation of the Skin Depth}
\label{subsec:derivation_skindepth}

We assume that the Sun's corona is magnetically dominated and ignore buoyancy and
gas pressure. If the fluid exterior to the loop is uniformly magnetized and possesses
a spatially constant  density, the MHD equations for the fluid velocity and magnetic
pressure fluctuation become particularly simple. The magnetic pressure fluctuation,
$\Pi$, obeys a standard wave equation \citep[e.g.,][]{Diaz:2003,Roberts:2019},

\begin{equation}
	\label{eqn:PDE_Pi}
	\dparn[\Pi]{t}{2} = V_e^2 \, \nabla^2 \Pi \; ,
\end{equation}

\noindent and the fluid velocity $\bvec{u}$ simultaneously obeys

\begin{equation}
	\label{eqn:PDE_u}
	\dparn[\bvec{u}]{t}{2}-V_e^2 \dparn[\bvec{u}]{z}{2} = -\frac{1}{\rho_e} \grad_\perp \dpar[\Pi]{t} \; ,
\end{equation}

\noindent where $V_e$ and $\rho_e$ are the Alfv\'en speed and mass density of the
external fluid, respectively. The spatial coordinate $z$ is aligned with the magnetic
field and the differential operator $\grad_\perp$ is the gradient in the plane
perpendicular to the magnetic field. These equations describe only Alfv\'en waves
and fast MHD waves. The slow magnetoacoustic waves are missing, because, as already
stated, we have adopted a ``cold-plasma" limit where the gas pressure is ignored.
Alfv\'en waves arise from the condition $\Pi=0$ and fast waves occur when the magnetic
pressure is nonzero.

From Equation~\eqnref{eqn:PDE_Pi}, we can directly extract an expression for the
evanescence length lateral to the field lines. By assuming exponential solutions
of the form,

\begin{equation}
	\Pi = e^{-i\omega t} \, e^{i k_z z} \, \tilde{\Pi}\left(\bvec{x}_\perp\right) \; ,
\end{equation}

\noindent we obtain an equation for the transverse variation of the magnetic pressure,

\begin{equation} \label{eqn:PDE_perp}
	\nabla_\perp^2 \Pi = \left(k_z^2 - \frac{\omega^2}{V_e^2}\right) \, .
\end{equation}

\noindent In these last two equations, $\omega$ is the temporal frequency, $k_z$
is the longitudinal wavenumber, and $\bvec{x}_\perp$ is the component of the 
position vector that is transverse to the magnetic field.

Up to this point in our derivation, we have implicitly assumed that the curvature
of the magnetic field lines can be ignored, but we have not assumed anything about
the shape or geometry of the loop's cross-section. From Equation~\eqnref{eqn:PDE_perp}
it is clear that the specific shape of the cross-section does not play much of a
role since the Laplacian operator is isotropic. For example, if we assume that the
fast waves are evanescent in the external fluid, and adopt an exponential form
that is valid in Cartesian geometry,

\begin{equation}
	\Pi = e^{-i\omega t} \, e^{i k_z z} \, \exp\left(-\bvec{\alpha} \cdot \bvec{x}_\perp\right) \; ,
\end{equation}

\noindent we obtain the following dispersion relation

\begin{equation}
	\omega^2 = \left(k_z^2 - \alpha^2\right) V_e^2 \, .
\end{equation}

\noindent The skin depth $\Delta$ is the reciprocal of the decay rate, $\Delta = 1/\alpha$.
Hence, the skin depth depends on the frequency, longitudinal wavenumber, and the
Alfv\'en speed,

\begin{equation} \label{eqn:SkinDepth}
	\frac{1}{\Delta^2} = k_z^2 - \frac{\omega^2}{V_e^2} \, .
\end{equation}

The same skin depth is achieved if we assume cylindrical symmetry \cite[e.g.,][]{Edwin:1983},

\begin{equation}
	\Pi = e^{-i\omega t} \, e^{i k_z z} \, e^{i \mu \phi} \, K_\mu\left(\alpha r\right) \; ,
\end{equation}

\noindent where $\phi$ is the azimuth, $r$ is the cylindrical radius, $\mu$ is the
azimuthal order, and $K_\mu$ is a modified Bessel function of the second kind. This
cylindrical form produces the same dispersion relation and skin depth.
The skin depth is independent of the transverse geometry. 

From Equation~\eqnref{eqn:SkinDepth}, we can extract a lower limit for the skin depth.
Clearly, the skin depth will be smallest (and the transverse decay rate the largest)
when the frequency is low (i.e., $\omega \ll k_z V_e$) and the second term on the
right-hand side can be ignored,

\begin{equation}
	\Delta > k_z^{-1} \, .
\end{equation}

\noindent We re-emphasize, that this lower limit is independent of the shape of the
loop's cross section and on the Alfv\'en speed. Further, its independent of direction.
For example, if one considers a loop with a rectangular cross-section, the external
wave solutions can have different decay lengths in the two transverse directions
\citep[e.g.,][]{Diaz:2003, Arregui:2007}.  But, increased lateral confinement in one
direction leads to decreased lateral confinement in the other because the decay rates
in the two directions add in quadrature,

\begin{equation}
	\frac{1}{\Delta^2} = \alpha^2 = \alpha_x^2 + \alpha_y^2 = \frac{1}{\Delta_x^2} + \frac{1}{\Delta_y^2} \; .
\end{equation}

\noindent In the previous expression, $\alpha_x$ and $\alpha_y$ are the two transverse
components of vector $\bvec{\alpha}$.

We can find a more nuanced expression for the skin depth if we assume that the
loop is thin and has a circular cross section. With these two assumptions we can
replace the frequency with the well-known dispersion relation for kink waves
on a slender magnetic flux tube, $\omega = k_z c_k$, where $c_k$ is given by
Equation~\eqnref{eqn:kink_speed} \citep[e.g.,][]{Edwin:1983}. Continuity of the
total pressure under the cold-plasma approximation dictates that the magnetic 
field has the same strength inside and outside the loop, $B_0 = B_e$. Hence, the
kink speed can be expressed in terms of the external Alfv\'en speed and the
overdensity ratio of the loop, $f \equiv \rho_0 / \rho_e$,

\begin{equation}
	c_k^2 = \frac{B_e^2}{2\pi \left(\rho_0 + \rho_e\right)} = \frac{2V_e^2}{f + 1}\, .
\end{equation}

Substituting this kink speed into the expression for the skin depth,
Equation~\eqnref{eqn:SkinDepth}, produces a frequency-independent result,

\begin{equation}
	\Delta = \left(\frac{f+1}{f-1}\right)^{1/2} k_z^{-1}\, .
\end{equation}

\noindent The leading coefficient involving the square root is always greater
than one; hence, the same lower limit that we derived earlier applies, $\Delta > 1/k_z$.
Here, we now see that this lower limit is approached as the loop becomes extremely
dense compared to the surrounding fluid, $\rho_0 \gg \rho_e$ or $f\gg1$. In
practice, the high overdensity limit is usually valid. Typical ``warm" coronal
loops (temperatures less than 3 MK) often have an overdensity ratio of 1000
or more \citep[e.g.,][]{Winebarger:2003}.

Finally, if we assume that the kink wave is a standing-wave resonance arising
from reflections at the two foot points where the magnetic loop intersects the
photosphere, we can relate the longitudinal wavenumber $k_z$, and subsequently
the skin depth $\Delta$, to the length of the loop $L$, 

\begin{equation} \label{fig:skin_depth_round}
	k_z = \frac{m \pi}{L} \; ,  \qquad \Delta =\left(\frac{f+1}{f-1}\right)^{1/2} \frac{L}{m\pi} \; .
\end{equation}

\noindent In these expressions, $m$ is the longitudinal mode order, with $m=1$
corresponding to the fundamental mode which lacks nodes in the velocity
eigenfunction. Almost all observed coronal loop oscillations are thought
to be oscillating in the fundamental mode and hence we will focus our
attention on $m=1$.

Since the lower limit for the skin depth of the fundamental mode, $\Delta > L/\pi$,
depends only on the loop length (and not the Alfv\'en speed or wave frequency),
the limit is useful and easily applied.  Unfortunately, it is also rather stringent
because the skin depth is so large. For example, for a coronal loop whose axis
is semicircular, the skin depth equals the radius of curvature and, accordingly,
the loop's maximum height above the photosphere. The skin depth is also comparable
to the spatial extent of many coronal arcades. Consider a ``typical" coronal loop
with a length of 100 Mm and an overdensity ratio of 10; the skin depth would be
35 Mm. Loops with a more moderate overdensity can have a significantly longer skin
depth. For instance, a 100 Mm loop that is twice as dense as its surroundings ($f=2$)
would have a skin depth that is nearly 50\% longer, $\Delta = 55$ Mm.

\subsection{Is a Coronal Loop Isolated?}
\label{subsec:isolation}

If the coronal loop is to oscillate in isolation, there cannot be any nearby magnetic
or density structures that efficiently scatter or reflect MHD waves, such as another
coronal loop. Any such structure that is sufficiently close will couple to the coronal
loop through wave scattering in their overlapping near fields. A good rule of thumb
is that the loop will interact appreciably with any scattering structure that is within
one or two skin depths. For example, \cite{Luna:2008} and \cite{VanDoorsselaere:2008}
found that this coupling could shift the frequency of kink oscillations up or down
by 25\% or more if two identical tubes were close---see Figure 3 of \cite{Luna:2008}
and Figure 6 of \cite{VanDoorsselaere:2008}. Both of these studies characterized the
``closeness" of the tubes in terms of the ratio $d/a$, where $d$ is the separation
between the centers of the two tubes and $a$ is the radius of the tubes. A more direct
criterion is based on whether the two tubes suffer near-field coupling, which depends
solely on the ratio of the separation to the skin depth, $d/\Delta$. It is straightforward
to work out the skin depth for these studies by using Equation~\eqnref{fig:skin_depth_round}.
In Figure 3 of \cite{Luna:2008} and Figure 6 of \cite{VanDoorsselaere:2008}, the skin
depth is 3.5 times the tube radius $a$. Thus, in both of these studies, it is clear
that strong coupling occurs when $d/\Delta < 1$ and modest coupling persists for
separations of up to two skin depths.

\begin{figure*}
	\epsscale{0.5}
	\plotone{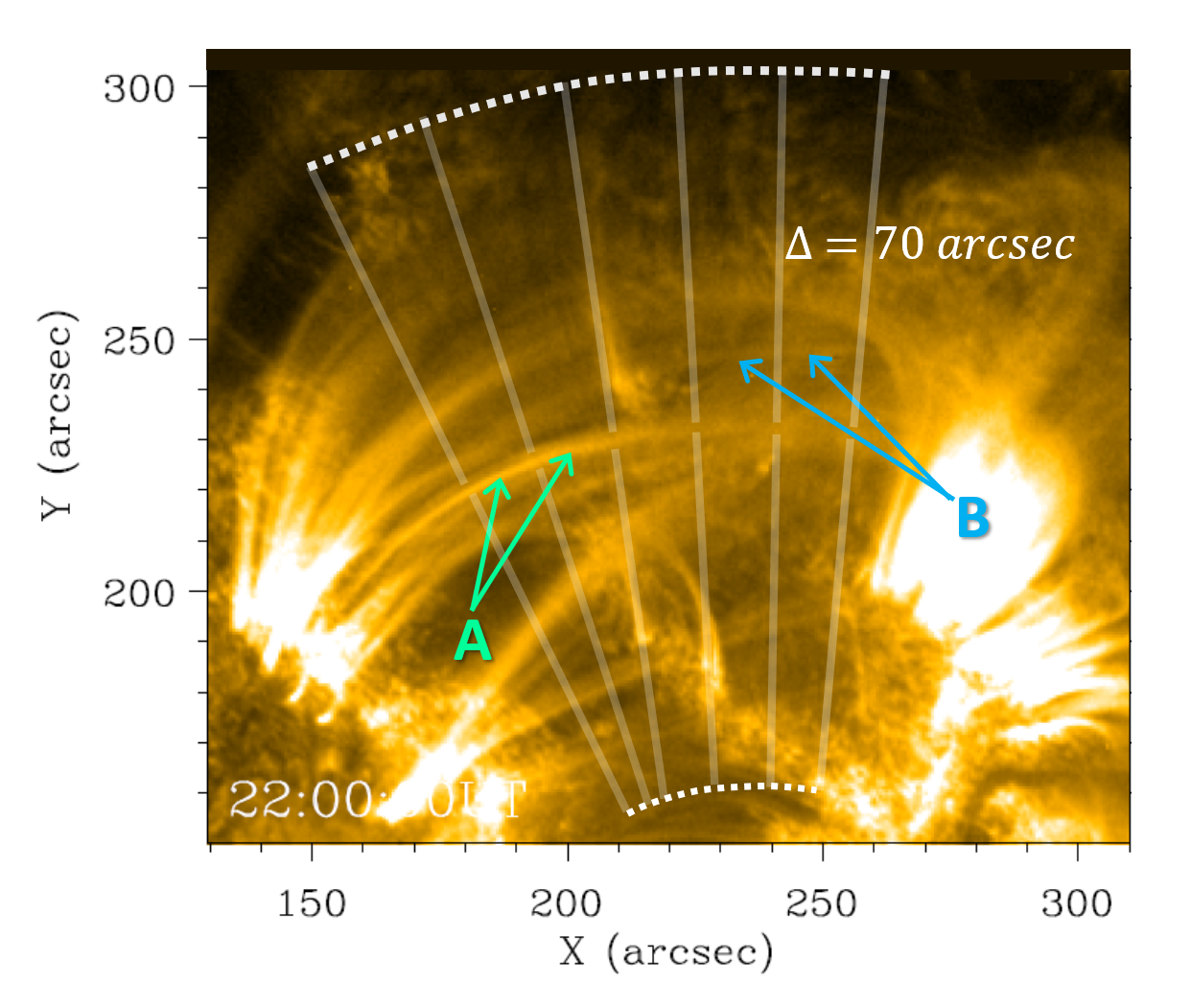}

	\caption{\small Snapshot of NOAA AR 1283 taken on 2011 September 6 by AIA
in the 171 \AA~ passband. There are a multitude of coronal loops evident, many
embedded in a common magnetic arcade. The two loops studied by \cite{Jain:2015}
are indicated and labeled A and B. Since each of these loops has a length of 160 Mm,
the lower limit for the skin depth of each is 50 Mm or 70 arcsec. Using the partially
transparent yellow lines, we have indicated a distance of 70 arcsec on the plane of
the sky to either side of loop A. The region between the dotted white lines indicates
the approximate region that lies within a single skin depth of loop A. Loops A and
B are clearly well within each others near field and hence, are strongly coupled.
In fact, most of field of view, and correspondingly much of the active region,
should be coupled to the oscillating loops.
	\label{fig:coupled_loops}}
\end{figure*}

For many coronal loops, satisfying the criteria that no other magnetic structure
lies within a skin depth is exceedingly difficult. In the catalog of loop oscillations
published by \cite{Nechaeva:2019}, 12 of the 20 loops have lengths exceeding 300 Mm.
Such loops will have skin depths in excess of 100 Mm.  It is difficult to believe
that the corona lacks scattering structures over such distances when active
regions themselves are ordered on a similar scale. There have been a variety of events
where coupling between nearby loops has been directly observed \citep[e.g.,][]{Schrijver:2000,
Schrijver:2002, Verwichte:2004, Jain:2015}. In \cite{Jain:2015} two loops with similar
length ($\sim$ 160 Mm) were seen to oscillate in concert. Figure~\ref{fig:coupled_loops}
shows an image of the two loops taken in the 171 \AA~ bandpass of the Atmospheric Imaging
Assembly (AIA; \cite{Lemen:2012}) aboard the {\it Solar Dynamics Observatory} ({\it SDO}).
\cite{Jain:2015} attributed the similarity in phase and initiation time of the two
loops to the near cotemporal excitation by a common driving event. However, considering that the loops
were only separated by 3.5 Mm (see Figure~\ref{fig:coupled_loops}), these loops were
deep within each other's near-field regions. Loops of the measured length, 160 Mm, will
have a skin depth exceeding 50 Mm. Since the loop separation is only 7\% of the skin
depth, it should be no surprise that the two loops were coupled and oscillated in synchronicity.
This would be true even if the excitation event (probably a wavefront launched by a nearby
flare) only acted to generate waves on one of the loops. Oscillation of one loop would
cause sympathetic vibration of the neighboring loop. In fact, far more than just the two
loops under consideration were coupled.  The skin depth of 50 Mm corresponds to roughly
70 arcsec. Most of the field of view in Figure~\ref{fig:coupled_loops} is within a single
skin depth of the two loops that were studied. Hence, to truly understand the oscillations
excited by the nearby flare, one needs to examine the entire active region, as all of the
magnetic features are coupled to some degree. This coupling of many magnetic structures
may also explain the slight phase shift that was observed between the two loops by \cite{Jain:2015}.
When only two loops are coupled, the collective modes involve either in-phase or
antiphase motions of the two tubes \cite[e.g.,][]{Luna:2008, Luna:2009, VanDoorsselaere:2009}.
But, such clear phase correlations are not expected when three or more oscillating
structures are coupled.

Even if a loop is isolated from other magnetic structures it's not free from the
tyranny of the skin depth. Most calculations and seismic analyses of the oscillation
frequencies assume that the curvature of the loop's axis can be ignored. However,
for a loop of length $L$ whose axis is semicircular, the radius of curvature of
the loop is identical to the skin depth, $L/\pi$. Thus, the loop should undergo
significant ``self-interaction" as the foot points of the loop lie within a skin
depth of each other. Of course, most loops will not form perfect semicircles.
Often, coronal loops are taller and with closer foot points than a semicircular
loop with identical length.  For such ``tall" loops, the foot points are closer than
a skin depth and the self-interaction will be even stronger. Even low-lying loops
with farspread  foot points are not immune. All portions of such a loop will lie
within a skin depth of the photosphere itself and we should expect significant
scattering from the photospheric surface that is not accounted for by the simple
line-tying boundary conditions imposed at the foot points.


\section{Propagating Fast Waves: Reflections from External Structures}
\label{sec:Propagating}

The coronal environment surrounding a coronal loop is usually an arabesque of
magnetic structures. Coronal loops are arranged in magnetic arcades like the warp
of a loom and these arcades are just one piece of a larger magnetic active region.
In the immediate aftermath of a nearby flare, waves are everywhere within the
active region. Fast MHD waves can be seen racing across field lines and often waves
can be observed to bounce back and forth across an arcade or the active region.
The animation shown as Figure~\ref{fig:movie} typifies such wave motion. Propagating
waves are seen to be ubiquitous and omnipresent.

Such behavior should be expected. The flare that excites oscillations on the coronal
loop is rarely if ever located on any of the field lines that pass through the loop.
Hence, fast MHD waves must travel from the flaring site, across field lines, to the
coronal loop where it excites kink waves. There is no reason to suppose that the
fast wave that performs this excitation does not excite oscillations on other magnetic
structures as it passes. These excited waves should bounce around the active region
until scattered, radiated, or dissipated.

The back and forth motion of the field lines within arcades suggests that the
arcade or active region may have resonances of its own, in addition to those of
coronal loops. Such arcade oscillations can clearly be seen to have significant
amplitude in coronal imagery (see Figure~\ref{fig:movie}). Hence, any potential
arcade resonances should have large enough amplitude to provoke substantial coupling
with the loop oscillations. Whether they do so or not will depends on whether the
arcade resonances have frequencies that are similar to the loop resonances. Here
we explore the possibility of coupling between the loop resonances and arcade
resonances using a simple model that is designed to be illustrative instead of predictive.


\subsection{Simple Model of an Arcade and Loop System}
\label{subsec:simple_model}

\begin{figure*}
	\epsscale{1.0}
	\plotone{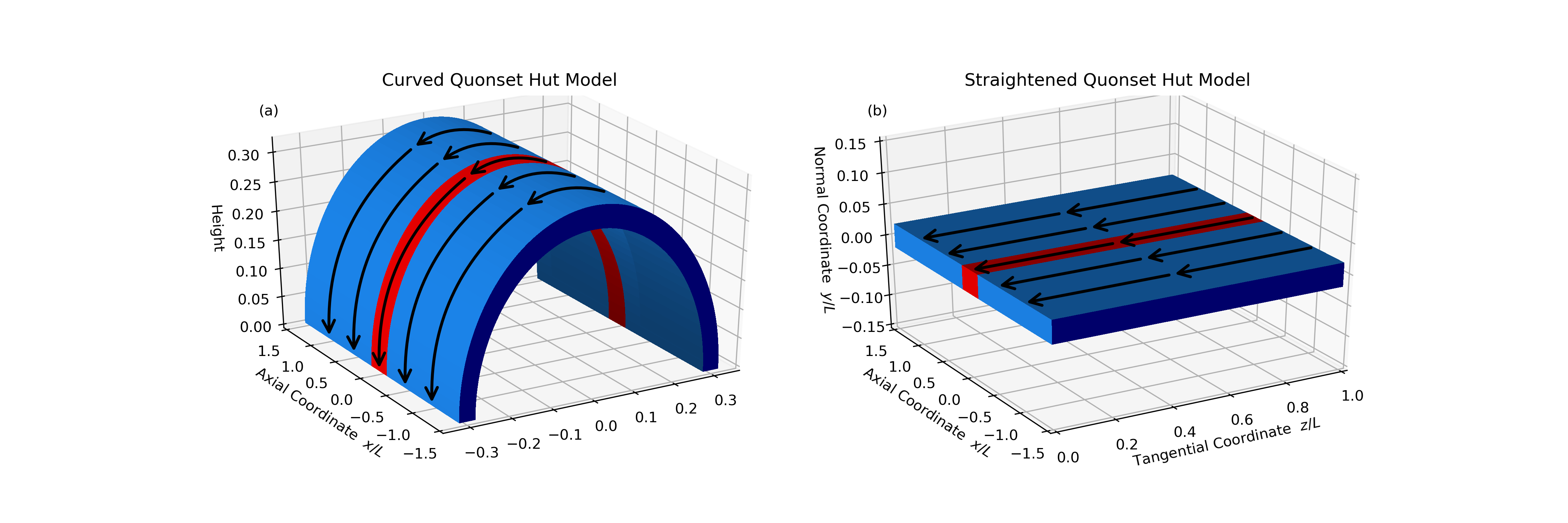}

	\caption{\small Geometry of our model of a coronal loop embedded in a coronal
arcade. ($a$) The arcade is a thin, curved sheet of plasma. The axis of the arcade
$\unitv{x}$, the magnetic field lines, and the curvature vector are all mutually
orthogonal. In the figure the arcade is indicated by the blue sheet and the magnetic
field lines by the black curves. Each field line intersects the photosphere twice. 
The loop is a segment of the arcade (shown in red), that lies in the center. ($b$)
We ignore the curvature of the arcade and loop, and work in slab geometry. The 
coordinate that is parallel to the magetic field is $z$. The coordinate parallel to
the axis of the arcade is $x$, and the coordinate normal to the sheet is $y$.
	\label{fig:geometry}}
\end{figure*}

We will model the arcade as a thin, curved, magnetic sheet with a thickness of $D$
and with field lines that pierce the photosphere twice. One can think of the arcade
as the roof of a Quonset hut and the photosphere as the ground. The two ribbons that
demark the intersection of the sheet with the photosphere are parallel to each other
and parallel to the arcade's geometrical axis, $\unitv{x}$ (see Figure~\ref{fig:geometry}).
The arcade's magnetic field $\bvec{B}$ lacks shear such that the field is everywhere 
perpendicular to arcade's axis, $\bvec{B} \cdot \unitv{x} = 0$. Therefore, for a
suitably thin sheet, we may assume that the length of each field line in the
corona---from photosphere to photosphere---is a constant value $L$. In the axial
direction, the magnetic sheet spans a finite width of $2W$ and the coronal loop
(shown in red in Figure~\ref{fig:geometry}) occupies the center of the arcade with
an axial extent of $2\delta$. The loop therefore has a rectangular cross section
with dimensions $2\delta \times D$ and a longitudinal length of $L$.

To keep the mathematics simple, we will make the standard assumption of ignoring
the curvature of the field lines in the MHD equations. This straightening of the
geometry permits us to utilize a coordinate system $\bvec{x} = (x,y,z)$ that is
coincident with the local Frenet coordinates for the field lines. The tangential
coordinate $z$ measures the distance along each field line, with $z=0$ and $z=L$
corresponding to the two points where the field line pierces the photosphere. The
coordinate $x$, points in the direction of the field line's binormal and measures
distances along the axis of the arcade. We place the origin, $x=0$, in the middle
of the loop and arcade system; hence, the arcade spans the domain $x\in[-W,W]$ and
the loop spans $x\in[-\delta,\delta]$. The third coordinate, $y$, is antiparallel
to the direction of the principle curvature of the field line, and points normal
to the magnetic sheet that forms the arcade. The geometry of the loop-arcade system
is a magnetized slab (loop) embedded within a less dense magnetized slab
(arcade)---see Figure~\ref{fig:geometry}.

As in the derivation of the skin depth, we will assume that the corona is magnetically
dominated and hence ignore gravity and gas pressure. Therefore, to enforce stability
in this straightened geometry, we must assume that the magnetic pressure is uniform.
This is most easily accomplished by adopting a constant magnetic field $\bvec{B} = B_0 \unitv{z}$.
To ensure that waves can be trapped in the coronal loop, we will consider a piece-wise
constant mass density and Alfv\'en speed, where the density in the arcade is $\rho_e$
and the density in the loop is $\rho_0$. The corresponding Alfv\'en speeds are $V_e$
and $V_0$. We assume that the loop is overdense compared to the surrounding arcade
$\rho_0 > \rho_e$ and, correspondingly, the Alfv\'en speed is reduced in the loop,
$V_0 < V_e$.


\subsubsection{Plane-Wave Solutions}
\label{subsubsec:planewaves}

Since, we have adopted a low-$\beta$ approximation and the Alfv\'en speed is
uniform within the loop and separately within the arcade, the waves are described
by Equations~\eqnref{eqn:PDE_Pi} and \eqnref{eqn:PDE_u} with the appropriate
Alfv\'en speed and density for each region. Since, we have ignored gas pressure
and buoyancy, the motions are purely transverse (i.e., $\bvec{u} = u_x \unitv{x} + u_y \unitv{y}$).
The solutions to Equations~\eqnref{eqn:PDE_Pi} and \eqnref{eqn:PDE_u} are plane
waves with frequency $\omega$ and wavenumber
$\bvec{k} = k_x \unitv{x} + k_y \unitv{y} + k_z \unitv{z}$, 

\begin{eqnarray}
	\Pi &\sim& \exp(-i\omega t) \, \exp(i \bvec{k} \cdot \bvec{x})  \, ,
\\
	\label{eqn:velocity_planewave}
	\bvec{u} &=& \frac{\omega \bvec{k}_\perp}{\rho\Omega^2} \Pi \, ,
\end{eqnarray}

\noindent where we have defined $\Omega^2 \equiv \omega^2 - k_z^2 V^2$ and the
frequency and wavenumber are related through a local dispersion relation

\begin{equation} \label{eqn:local_disp}
	\omega^2 = \left(k_x^2 + k_y^2 + k_z^2\right) V^2 \; .
\end{equation}

\noindent In these expressions the Alfv\'en speed, $V$ takes on the value of either
$V_0$ or $V_e$ depending on whether we are considering the region inside or outside
the loop. The wavenumber $\bvec{k}_\perp$ is the component of the wavenumber $\bvec{k}$
that is transverse to the magnetic field, i.e., $\bvec{k}_\perp = k_x \unitv{x} + k_y \unitv{y}$.
For those waves that are laterally evanescent, the transverse wavenumber is related
to the lateral decay rate, $\bvec{k}_\perp = i \bvec{\alpha}$.


\subsubsection{Boundary and Interface Conditions}
\label{subsubsec:boundary_conditions}

We apply boundary conditions, two in each spatial dimension, that are chosen largely
for convenience and to develop a simple, easily understood example. For the boundary
conditions in the longitudinal coordinate $z$, we recognize that the photosphere
can be treated as an extremely dense immovable plasma.  Hence, we adopt the standard
line-tying condition that the velocity must vanish at the two foot points of the
field line, i.e., $\bvec{u} = 0$ at $z=0$ and $z=L$. This requirement quantitizes
the tangential wavenumber, mandating that $k_z = m \pi / L$, with $m$ being a positive
integer. The longitudinal wavenumber will be the same in both regions, inside and
outside the loop.

In the direction normal to the arcade's sheet, $y$, we adopt impenetrable boundaries
at the top and bottom surfaces of the sheet, $y = \pm D/2$, and require that the
fluid motion is unvarying across the sheet, $k_y = 0$. We fully recognize that the
boundary condition of impenetability is a rather poor one; but without it, motions
in the two transverse directions become inherently coupled and a tremendous degree
of complexity ensues \citep[see for example][]{Joarder:1992}. By imposing impenetrable boundaries
and spatial invariance in the $y$ direction, we ensure that motions are allowed only
in the axial direction $x$ and correspond to so called `horizontal' oscillations of
the loop. 

For the third set of boundary conditions, we adopt ``open boundaries" at the ends of
the arcade in the axial direction, i.e., we require that magnetic pressure fluctuation
vanish at $x = \pm W$. A quick examination of the PDEs, Equations~\eqnref{eqn:PDE_Pi}
and \eqnref{eqn:PDE_u}, reveals that this condition is equivalent to requiring that
the axial derivative of the axial velocity vanishes at both end points, $\partial u_x /\partial x=0$.
Other homogeneous boundary conditions in the axial direction could be implemented.
For example, we could enforce impenetrable boundaries $u_x =0$, or we could
match to an extremely diffuse coronal background and require
evanescent solutions that decay away from the arcade. However, doing so does not
fundamentally change the structure of the eigenfunctions or the eigenfrequencies,
but can involve a large degree of added complexity.

With these choices of boundary conditions, the magnetic pressure and the velocity
must have the following functional form,

\begin{eqnarray}
	\Pi &=& \sin\left(\frac{m \pi}{L} z\right) \, e^{-i\omega t} \nonumber
\\ && \; \times
		\left\{
		\begin{array}{ll}
			A_1 \cos\left(k_x x\right) + A_2 \sin\left(k_x x\right) & {\rm ~for}~|x|<\delta \\
			A_3 \sin\left[k_x(W-x)\right] &{\rm~for}~|x|>\delta
		\end{array}
		\right.
\\ \nonumber \\
	u_x &=& \frac{i \omega k_x}{\rho\Omega^2} \sin\left(\frac{m \pi}{L} z\right) \, e^{-i\omega t}  \nonumber
\\ && \; \times
			\left\{
			\begin{array}{ll}
				A_1 \sin\left(k_x x\right) - A_2 \cos\left(k_x x\right) & {\rm~for}~|x|<\delta \\
				A_3 \cos\left[k_x(W-x)\right] & {\rm~for}~|x|>\delta
			\end{array}
			\right.
\end{eqnarray}

\noindent with $A_1$, $A_2$, and $A_3$ being real-valued amplitudes.

The final requirements that are needed to fully specify the solution are matching conditions
that apply at the interfaces between the loop and arcade, $x=\pm \delta$. We require
that both the magnetic pressure and the axial velocity are continuous across these
interfaces. The simultaneous solution of these four interface constraints has a solvability
condition that results in a global, transcendental dispersion relation. The modes
that are odd functions about the center of the loop, $x=0$, have a different dispersion
relation than those that are even. The solutions that have an even velocity eigenfunction
(or odd magnetic pressure fluctuation) are called kink waves and those that have an
odd velocity function (or even magnetic pressure fluctuation) are called sausage waves
\citep{Edwin:1982}.
It is a confusing matter of nomenclature, that both these slab modes, kink and sausage,
are fast MHD waves (remember we have no slow waves because $\beta=0$). In cylindrical
geometry, for a flux tube with low but finite $\beta$, the kink wave is also a type
of fast wave but the sausage wave is a varicose slow MHD wave.

The dispersion relations of the two waves modes are as follows,

\begin{eqnarray} \label{eqn:disp_kink}
	{\rm kink:} \;\; k_e \tan\left[k_e(W-\delta)\right] + k_0 \tan\left(k_0\delta\right) &=& 0 \; ,
\\ \label{eqn:disp_saus}
	{\rm sausage:} \;\; k_e \tan\left[k_e(W-\delta)\right] - k_0 \cot\left(k_0\delta\right) &=& 0 \; ,
\end{eqnarray}

\noindent where $k_e$ and $k_0$ are the axial wavenumbers in the arcade and in the
loop, respectively,

\begin{eqnarray}
	k_e^2 = \frac{\omega^2}{V_e^2} - \left(\frac{m\pi}{L}\right)^2 = \frac{\Omega_e^2}{V_e^2} \; ,
\\
	k_0^2 = \frac{\omega^2}{V_0^2} - \left(\frac{m\pi}{L}\right)^2 = \frac{\Omega_0^2}{V_0^2} \; .
\end{eqnarray}

\noindent Note, each region has a distinct cutoff frequency, $\omega_c = m\pi L / V$,
which depends on the loop length $L$ and the appropriate Alfv\'en speed for the
region under consideration, $V = V_e$ or $V=V_0$. For frequencies above the cutoff,
the waves are axially propagating, $k_e^2 > 0$ or $k_0^2 > 0$.


\subsubsection{Eigenfrequencies and Eigenfunctions}
\label{subsubsec:eigenstuff}

The dispersion relations can be solved numerically for the wave frequency as a
function of the two Alfv\'en speeds, $V_0$ and $V_e$, the loop length $L$, the
half-width of the loop $\delta$, the breadth of the arcade $2W$, and the longitudinal
mode order $m$. In order to reduce the number of free parameters that the frequency
depends on, we consider only the longitudinal fundamental $m=1$, and we use the
loop length $L$ and the Alfv\'en crossing time $L/V_0$ to nondimensionalize. With
this change of variable, the dispersion relation can be solved for the nondimensional
frequency, $\varpi \equiv (L/V_0) \omega$, in terms of three ratios: the
nondimensional half width of the loop $\delta/L$, the nondimensional breadth of the
arcade $2W/L$, and the overdensity ratio $f=\rho_0 / \rho_e = V_e^2 / V_0^2$.

Once an eigenfrequency has been obtained through the transcendental dispersion
relation, the magnetic pressure and velocity eigenfunctions that are associated
with that frequency are given by the following equations,

\vspace{0.2in}
{\it Kink Modes}
\begin{eqnarray}
		\Pi &=& \frac{B_0^2}{4\pi} \sin\left(\frac{m \pi}{L} z\right) \, e^{-i\omega t}	
\\ 		&& \; \times \left\{
		\begin{array}{ll}
			\sin(k_0 x) & {\rm~for~} |x|<\delta \\
			\D \frac{{\rm sgn}(x) \sin\left(k_0 \delta\right)}{\sin\left[k_e \left(W-\delta\right)\right]} \sin\left[k_e\left(W-|x|\right)\right] &  {\rm~for~} |x|>\delta
		\end{array}
		\right.
\nonumber\label{eqn:EigKink}
\\ \nonumber \\ \nonumber \\
		u_x &=& i \frac{\omega}{k_0} \sin\left(\frac{m \pi}{L} z\right) \, e^{-i\omega t}
\\ 		&& \; \times \left\{
		\begin{array}{ll}
			-\cos(k_0 x) & {\rm~for~} |x|<\delta \\
			\D \frac{k_0 \sin\left(k_0 \delta\right)}{k_e \sin\left[k_e \left(W-\delta\right)\right]} \cos\left[k_e\left(W-|x|\right)\right] &  {\rm~for~} |x|>\delta
		\end{array}
		\right.
\nonumber		
\end{eqnarray}

\vspace{0.2in}
{\it Sausage Modes}
\begin{eqnarray}\label{eqn:EigSausage}
		\Pi &=& \frac{B_0^2}{4\pi} \sin\left(\frac{m \pi}{L} z\right) \, e^{-i\omega t}
\\		&& \; \times \left\{
		\begin{array}{ll}
			\cos(k_0 x) & {\rm for~} |x|<\delta \\
			\D \frac{\cos\left(k_0 \delta\right)}{\sin\left[k_e \left(W-\delta\right)\right]} \sin\left[k_e\left(W-|x|\right)\right] &  {\rm for~} |x|>\delta
		\end{array}
		\right.
\nonumber \\ \nonumber \\
		u_x &=& i \frac{\omega}{k_0} \sin\left(\frac{m \pi}{L} z\right) \, e^{-i\omega t} 
\\		&& \times \left\{
		\begin{array}{ll}
			\sin(k_0 x) & {\rm if~} |x|<\delta \\
			\D \frac{{\rm sgn}(x) k_0 \cos\left(k_0 \delta\right)}{k_e \sin\left[k_e \left(W-\delta\right)\right]} \cos\left[k_e\left(W-|x|\right)\right] &  {\rm if~} |x|>\delta
		\end{array}
		\right.
\nonumber	
\end{eqnarray}

\begin{figure*}
	\epsscale{0.5}
	\plotone{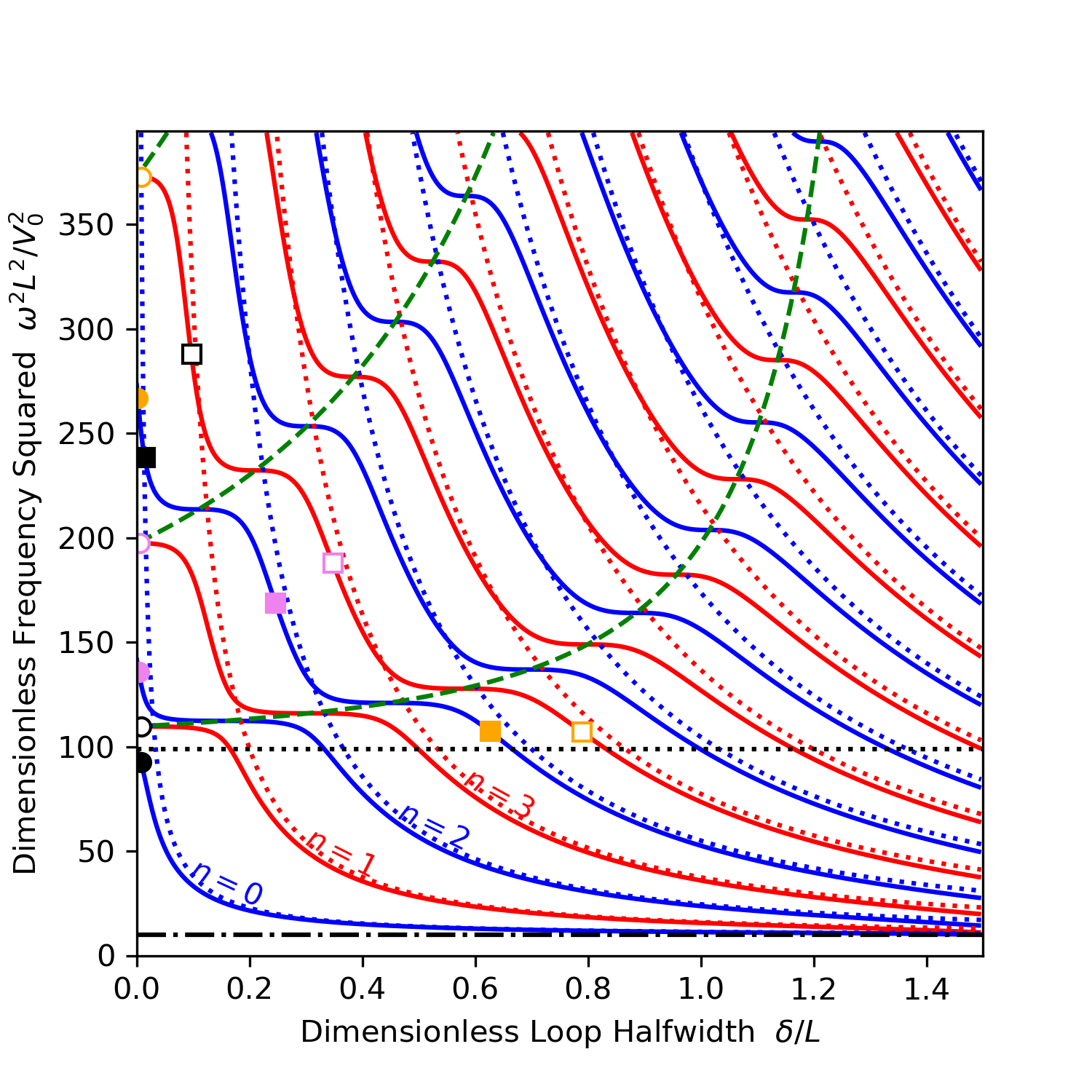}

	\caption{\small Mode diagram that indicates the frequency of modes as a
function of the loop's half width. The blue curves correspond to kink modes and the
red curves to sausage modes.  All modes correspond to the longitudinal fundamental,
$m=1$.  The solid curves indicate the modes of the loop-arcade system and the dotted
curves show the resonances of the loop in isolation, in the limit of large loop
overdensity ($f\gg1$) and extremely wide arcade ($W\gg\delta$).  The axial order,
$n$, of each modal curve is indicated for the first four overtones. The dotted horizontal
line indicates the cutoff frequency for the arcade, $m\pi L / V_e$, and the dot-dashed
line indicates the cutoff for the loop, $m\pi L / V_0$.  Waves with frequencies above
the cutoff are laterally propagating within the associated region. The dashed green
curves indicate the modes of the arcade and correspond to where a node in the axial
velocity is coincident with the boundary between the arcade and loop. For each dispersion
curve, as the width of the loop decreases, the frequency increases. Whenever a node moves
from inside the loop to outside (and the dispersion relation crosses a green dashed curve),
an avoided crossing occurs. These crossings can be thought of as the parameter values
for which a mode of the arcade has the same frequency as a mode of the loop. The
colored squares and circles indicate modes that have eigenfunctions that are illustrated
in subsequent figures. The filled squares are kink modes with four total axial nodes
in the velocity eigenfunction. The open squares are sausage modes with five nodes.
The eigenfunctions are illustrated in Figure~\ref{fig:eigenfunctions}. The circles
correspond to kink modes (filled) and sausage modes (open) in the thin-loop limit,
and their eigenfunctions are shown in Figure~\ref{fig:thinslab_eigfunctions}.
	\label{fig:dispersion}}
\end{figure*}

For any set of parameter values, there are of course a countable infinity of eigenfrequencies.
Figure~\ref{fig:dispersion} is a mode diagram that illustrates the mode frequencies
as a function of the loop's half-width for an overdensity ratio of $f=10$ and an
arcade breadth of $2W/L=3$. Only the longitudinal fundamental, $m=1$, is illustrated.
The blue curves indicate the frequency of kink modes
and the red curves show sausage modes. We label each member of the sequence by the
axial mode order $n$. This order is a non-negative integer that indicates the number
of nodes that appear in the axial direction for the velocity eigenfunction, $u_x$.
The gravest mode, $n=0$, corresponds to the traditional kink oscillation of a thin
coronal loop. It lacks nodes in the velocity eigenfunction, has a single pressure
node in the middle of the loop, and is evanescent in the region outside the loop. In
Figure~\ref{fig:dispersion}, the two horizontal black lines indicate cutoff frequencies
in the loop (dash-dotted line), $m \pi V_0/L$, and in the arcade (dotted line),
$m \pi V_e/L$. Waves with frequencies less than the respective cutoff are axially
evanescent ($k_x^2 < 0$) in the associated region.

The higher orders $n>0$, correspond to kink and sausage modes with additional axial
nodes, i.e., axial overtones. For thin loops (small $\delta$) all of the overtones
propagate axially in the arcade, $k_e^2 > 0$.  Only for sufficiently fat loops does
the frequency of an overtone drop below the cutoff frequency for the arcade and become
evanescent outside the loop. The threshold loop thickness varies from overtone to overtone.

A prominent feature of the dispersion curves for all of the overtones are a sequence
of avoided crossings. These crossings occur when a node in the velocity eigenfunction
passes across the loop-arcade boundary and can be thought of as those parameter values
where a resonance of the loop and a resonance of the arcade have similar frequencies.
If the arcade were to be very wide with extremely low density (high
Alfv\'en speed), the modes of the loop would be given by a transcendental dispersion
relation \citep{Edwin:1982},

\begin{eqnarray}
	{\rm kink:} \qquad -|k_e| + k_0 \tan\left(k_0\delta\right) &=& 0 \; ,
\\
	{\rm sausage:} \qquad -|k_e| - k_0 \cot\left(k_0\delta\right) &=& 0 \; .
\end{eqnarray}

\noindent These equations can be derived from Equations~\eqnref{eqn:disp_kink} and
\eqnref{eqn:disp_saus}, by assuming that $f\gg1$ and $W\gg\delta$.  The resulting
sausage waves and kink waves are shown as dotted red and blue curves in
Figure~\ref{fig:dispersion}. In the same limit of a very dense loop ($f\gg1$), the
modes of the arcade correspond to those waves that treat the loop as immovable,
i.e., those that have $u_x=0$ at the loop-arcade boundaries, $x=\pm\delta$.  One can
easily derive from Equations~\eqnref{eqn:EigKink} and \eqnref{eqn:EigSausage} the
value of the axial wavenumber which generates a node at this interface,

\begin{equation}
	k_e = \frac{\left(j+1/2\right)\pi}{W-\delta} \qquad {\rm for~} j = 0, 1, 2, 3, \ldots \; .
\end{equation}

\noindent By using the local dispersion relation, Equation~\eqnref{eqn:local_disp}, this
condition on the wavenumber can be rewritten in terms of the wave frequency,

\begin{equation}
	\omega^2 = \left[\frac{m^2\pi^2}{L^2} + \frac{\left(j+1/2\right)^2\pi^2}{\left(W-\delta\right)^2}\right] V_e^2 \; .
\end{equation}

\noindent These frequencies are indicated on Figure~\ref{fig:dispersion} using the
dashed green curves with $j=0$ being the lowest curve and $j=1$ and $j=2$ appearing
at higher frequencies.  It is clear from the diagram that a mode of the loop-arcade
system can be roughly classified as either a loop mode or an arcade mode, except
where the frequencies of the two families of solutions are commensurate.  At these
points in parameter space, a mode of the system is mixed in character and a classic
avoided crossing occurs as the loop width parameter is varied.

\begin{figure*}
	\epsscale{1.0}
	\plotone{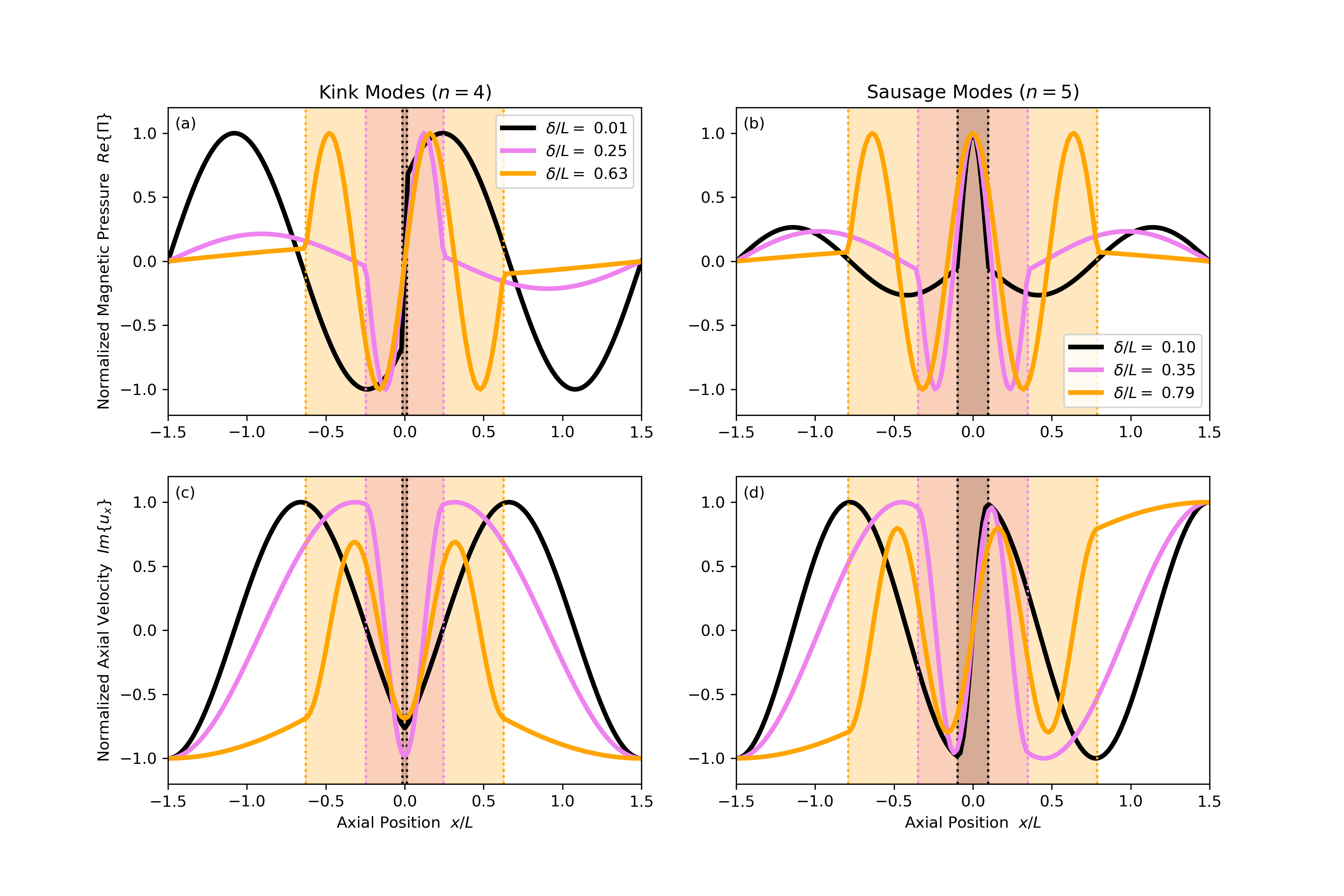}

	\caption{\small Eigenfunctions for a selection of kink modes (left panels)
and sausage modes (right panels). The upper panels show the magnetic pressure
fluctuation and the bottom panels present the axial velocity. All of the kink
modes lie on the $n=4$ dispersion curve in Figure~\ref{fig:dispersion} and
are indicated by the filled squares in that figure. Accordingly all of these
kink modes have four nodes in their eigenfunctions for the axial velocity.
Similarly, all of the sausage modes have five nodes, lie on the $n=5$ dispersion
curve, and are indicated by the open squares. The color of each curve indicates
which mode is being plotted, with the same color used to label the mode in
Figure~\ref{fig:dispersion}. Since, the modes are for different values of the
loop's half-width, $\delta$, the region occupied by the loop is indicated by
the color of the shaded regions.  The yellow color indicates the widest loop,
and taupe the narrowest.
	\label{fig:eigenfunctions}}
\end{figure*}

The change in behavior as a node slides through the loop boundary is further
illustrated in Figure~\ref{fig:eigenfunctions}, which shows eigenfunctions
for a $n=4$ kink mode (left panels) and a $n=5$ sausage mode (right panels). The
top panels show the magnetic pressure fluctuation and the bottom panels are the
axial velocity. Since the axial velocity is purely imaginary when the magnetic
pressure fluctuation is purely real, we show the imaginary part of the axial velocity.
All eigenfunctions have been normalized to have a peak value of unity.

The three different kink-mode curves correspond to different values of the loop's
half-width $\delta$. The three values are printed in the legend, and are also
indicated with the filled squared in the mode diagram, Figure~\ref{fig:dispersion}.
The color of the squares in Figure~\ref{fig:dispersion} matches the color of the
eigenfunction curves in Figure~\ref{fig:eigenfunctions}. These three values of
the loop width were chosen to fall between avoided crossings in the mode diagram.
One can easily determine by inspection that all velocity eigenfunctions for the
kink mode have four nodes ($n=4$), but depending on where the mode falls on the
mode diagram in relation to the avoided crossings, there can be zero, two or four
nodes within the loop itself. A similar result holds for the sausage mode with
five total nodes and a differing number of nodes located within the loop. The
three values of the loop width that are illustrated for the sausage mode are
indicated with open, colored squares in the mode diagram.

\subsection{Are Loop Oscillations Due to a Local Resonance?}
\label{subsec:local_resonances}

The width of a coronal loop compared to its length is typically quite small. This
fact has been exploited theoretically, by considering the limit where the width
of the loop goes to zero. This so called ``thin loop" limit is exceedingly advantageous
as it permits the coronal seismologist to ignore any variations across the cross
section of the loop. It is not that the loop lacks substructure; the loop may indeed 
be comprised of a bundle of threads. Instead, the substructure does not matter seismically.
The threads oscillate collectively, and for seismic purposes the loop may be treated
as a narrow, featureless bundle of field lines \citep[e.g.,][]{Terradas:2008}.

The equivalent limit for our loop-arcade model is the thin slab limit, where we
consider what happens when we let the loop's half-width approach zero, $\delta / L \to 0$.
In this limit it is easy to demonstrate that the frequency of the fundamental kink
mode, $n=0$, converges to the well-known result $\omega \to k_z V_e$ \citep[i.e.,][]{Edwin:1982}.
That is, the longitudinal phase speed approaches the external Alfv\'en speed,
$\omega/k_z \to V_e$. This can be seen directly in Figure~\ref{fig:dispersion},
where the gravest mode approaches the arcade's cutoff frequency as $\delta \to 0$.
It is also simple to demonstrate that the overtones ($n>0$) all converge to the
eigenfrequencies that the arcade would have in the absence of the loop,

\begin{equation} \label{eqn:thin_arcade}
	\omega^2 \to \left(\frac{n^2\pi^2}{4W^2}+\frac{m^2\pi^2}{L^2}\right) V_e^2 \; .
\end{equation}

\noindent Thus, when the loop is thin, there is a clear separation
of the two families of modes: the fundamental mode ($n=0$) corresponds to a loop mode
that resides primarily in the loop and axial overtones ($n>0$) are arcade modes that
reside primarily in the arcade.  When the loop is thick (finite $\delta$), the situation
becomes more complicated.  At low frequency, all of the modes have an energy density
confined to the loop.  But, at higher frequencies, the modes propagate both inside
and outside the loop.   

\begin{figure*}
	\epsscale{1.0}
	\plotone{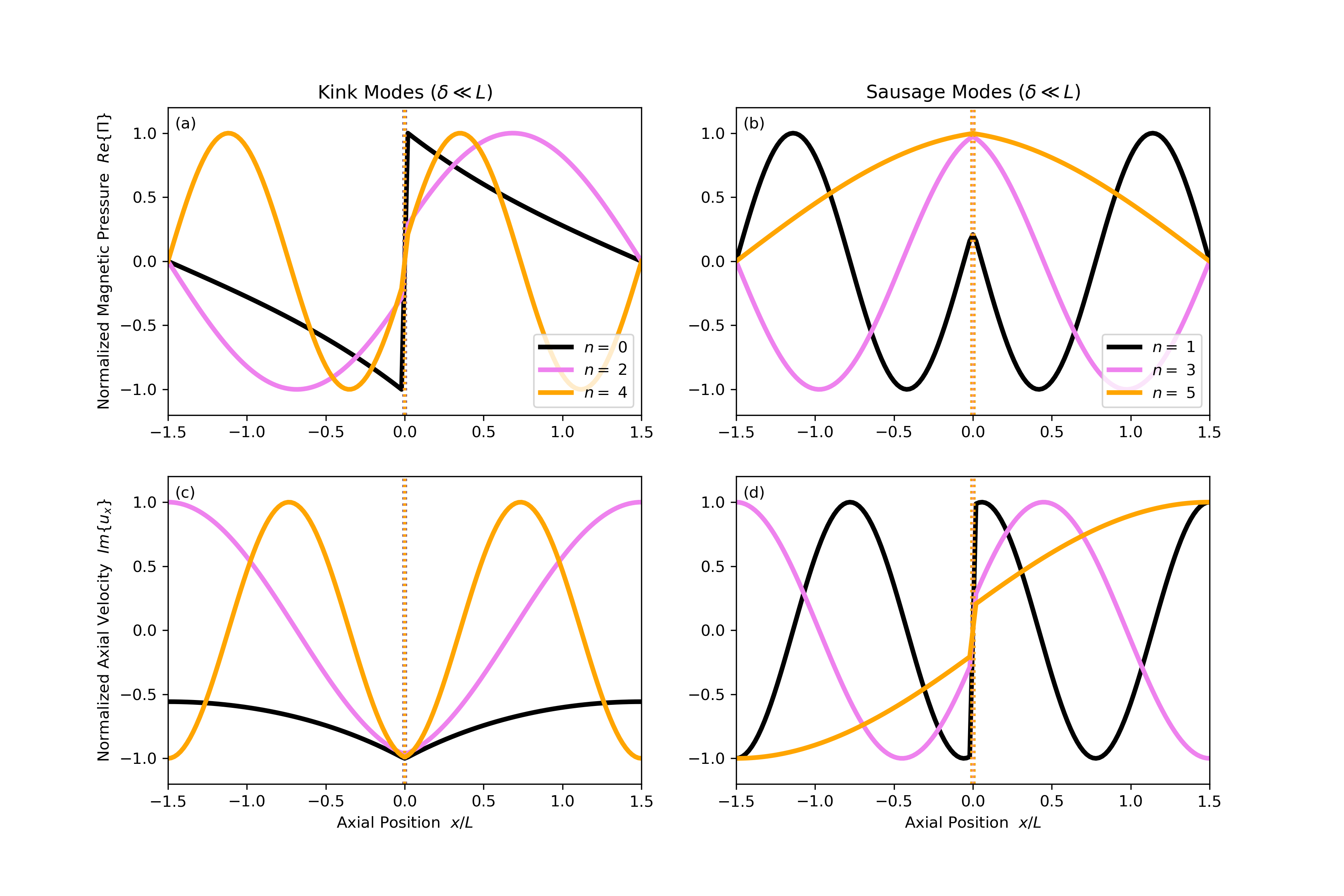}

	\caption{\small Eigenfunctions for a selection of kink modes (left panels)
and sausage modes (right panels) that are within the thin loop limit. The upper 
panels show the magnetic pressure fluctuation and the bottom panels present the 
axial velocity. The three kink modes correspond to three different axial orders,
$n= 0$, 2, and 4. The three modes that are illustrated are indicated on the 
mode diagram, Figure~\ref{fig:dispersion}, with filled circles.  The color
of the circle corresponds to the color of the eigenfunction curve.  Similarly,
the three sausage modes are for the first three odd orders, $n=1$, 3, and 5. They
are indicated in the mode diagram by the open circles. The $n=0$ is the only mode
that is laterally evanescent within the arcade and as such is a loop resonance.
All of the other modes propagate within the arcade and are arcade resonances.
	\label{fig:thinslab_eigfunctions}}
\end{figure*}

\begin{figure*}
	\epsscale{1.2}
	\plotone{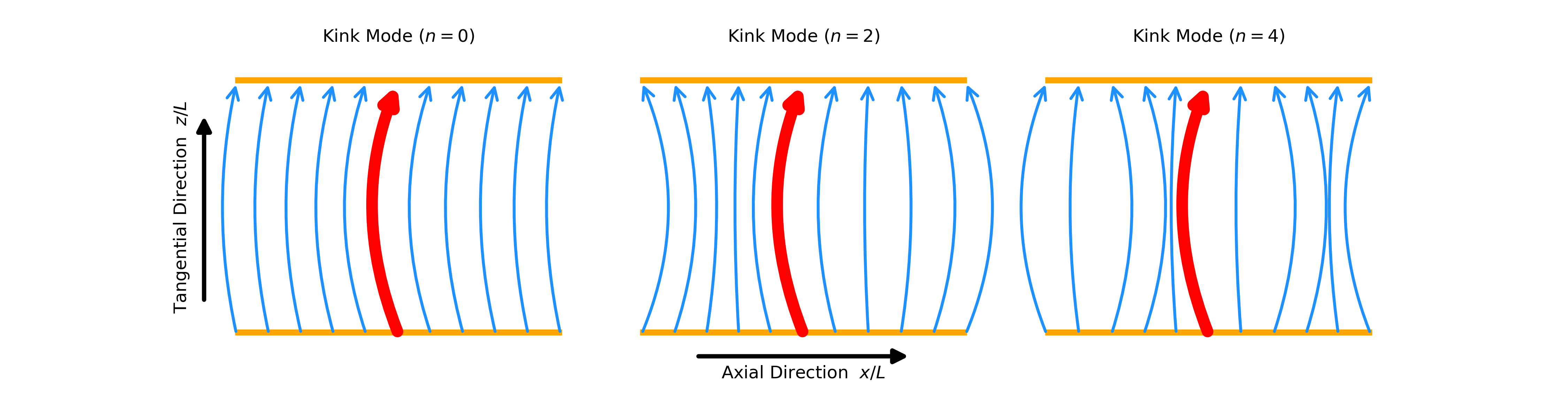}

	\caption{\small Snapshot of the displacement of field lines caused by three
different kink modes, $n=0$, 2, and 4. The gold lines indicate the foot points where
the field lines in the arcade intersect the photosphere. The red curved red arrow
shows the position of the coronal loop. Note, the displacement of the loop is same
for all three modes. The blue arrows indicate the displacement of field lines outside
the loop within the arcade.  All differences between the three modes occur within the
arcade. Often the motions within the host arcade are  invisible due to dimness and poor
spatial contrast. Thus, all three of these modes would look the same if all one can
see is the coronal loop. This can lead to mode misidentification and substantial error
in the seismic measurement of loop properties. 
	\label{fig:misidentification}}
\end{figure*}

At first blush, this separation of the modes into arcade modes and loop modes for
thin loops sounds promising.  It suggests that coronal seismology should
be able to concentrate on the loop resonance and ignore the detailed properties
of the coronal arcade. The initial glee resulting from this impression should
be tempered, however. Mode identification turns out to be problematic. Typically,
the swaying motion of a coronal loop is relatively easy to observe and measure because
the loop is bright. The associated motions within the surrounding arcade may be largely
invisible because the arcade is dim and smooth with little spatial variation in brightness.
Unfortunately, this means that all kink modes ($n=0, 2, 4, 6, \ldots$) will
look essentially the same, independent of whether they correspond to a loop resonance
or an arcade resonance. For all of these modes, the coronal loop will be seen to sway
back and forth because all modes lack structure or nodes within the loop itself.
Figure~\ref{fig:thinslab_eigfunctions} illustrates three of the gravest kink modes
$n=0$, 2, and 4 and sausage modes $n=1$, 3, and 5 in the thin slab limit. The three
kink modes are indicated on the mode diagram, Figure~\ref{fig:dispersion}, with
the filled colored circles. The sausage modes are marked with the open circles. The
color of the circle corresponds to the color of the eigenfunction curves. It is clear,
that within the loop, all of the kink modes have essentially the same eigenfunction
and separately so do all the sausage waves. Another view of this is provided in
Figure~\ref{fig:misidentification}, where a snap shot of the oscillating field lines
for the same three kink modes are shown. The blue lines correspond to field lines in
the arcade and the thick red line shows the position of the loop. Note, the displacement
amplitude of each of the illustrated field lines is calculated from the eigenfunctions,
Equation~\eqnref{eqn:EigKink}. As one can see, the shape of the loop looks identical
for all three modes. 

For the $n=0$ mode, the wave is a resonance of the loop and the loop is actively
participating in the oscillation. For the overtones, $n>0$, the loop is a bright
passive tracer that follows the motion within the surrounding arcade. If you do not
directly observe the motions within the arcade, the only observable distinction
between the resonance of the loop and resonances of the arcade is the frequency of
oscillation. But, the frequency is the quantity that one must measure to seismically
assess the properties of the loop.

Most large-amplitude coronal loop oscillations seem to be connected to flares,
but the mechanism is not well understood. Often a wave front is seen propagating
away from the flare site \citep{Jain:2015, Zhang:2020} and such fronts may be coeval with
the initiation of the oscillations. Such a mechanism should indiscrimately excite
a broad spectrum of arcade and loop modes.  In fact, its hard to imagine how only
the loop resonance could be selectively excited while the arcade modes remain dormant.
In fact, it seems more likely that the converse would occur: a wavefront launched
by a flare impacts the edges of the arcade and preferentially excites the gravest
arcade modes, only weakly tickling the loop resonance because the loop is embedded
in the arcade.

Misidentification of a mode has unfortunate consequences. How much of an error in
a seismic inference is likely?  In general the ratio of mode frequencies of an
axial overtone to the axial fundamental in the thin loop limit is given by,

\begin{equation} \label{eqn:freq_ratio}
        \frac{\omega_n}{\omega_0} = \left(1 + \frac{n^2 L^2}{4m^2 W^2}\right)^{1/2} \; .
\end{equation}

\noindent Imagine that a flare excites the gravest kink mode of the arcade ($n=2$),
but one mistakenly believes that the oscillation is just of the embedded loop ($n=0$).
For a thin loop, the ratio of the frequencies of the two modes is given by

\begin{equation}
	\frac{\omega_2}{\omega_0} = \left(1 + \frac{L^2}{W^2}\right)^{1/2} \; ,
\end{equation}

\noindent where we have assumed $m=1$. Note, the frequency ratio only depends
on the physical dimensions of the arcade. Since, the seismic estimate of the
kink speed is proportional to the wave frequency, the fractional error that is
made in the inferred kink speed is equal to the fractional difference between
the two mode frequencies. If we assume a loop length of $L=100$ Mm and an arcade
breadth of $2W = 200$ Mm, the resulting fractional error is around a 40\%
overestimation of the wave speed (and hence the magnetic pressure).


\section{Discussion}
\label{sec:Discussion}

The goal of coronal seismology is to probe the density and magnetic field strength
of a corona loop (and potentially its coronal environment) through the measurement
of the resonant oscillation frequencies of the loop. The seismic inferences are
therefore highly dependent on the validity of the wave model that predicts the
frequencies as functions of the properties of the loop. If a poor model is employed, the
seismic deductions will be equally poor.  To date, the models used in coronal
seismology have been predicated on the axiom that the observed oscillations are
resonant kink oscillations of an isolated magnetic loop. In general, the effects
of structure in the loop's environment have received only cursory attention. Here
we examine two ways in which this axiom can be broken: (1) we question the isolation
of coronal loops from other magnetic structures and (2) we explore potential interactions
of the loop resonances with the resonances of the magnetic arcade that cradles the loop.


\subsection{Coupling with Nearby Structures}
\label{subsec:coupling_structure}

Coronal loops are always part of a larger flux system or arcade and these arcades
tend to be highly structured.  Often many loops can be observed within the same
arcade, sometimes visible and sometimes not, as each loop brightens and fades.
{\it In order for the measured frequencies to be a direct diagnostic of the properties
of the vacillating loop, the arcade that houses the loop must be weakly structured
such that no significant scatterer lies within a skin depth.} Since, the skin depth
of a slender magnetic loop is generally quite large, often comparable to the height
of the loop itself, the various loops within an arcade are often within a skin depth
of each other, as are the edges of the arcade. Thus, the oscillations of most observed
loops are the response of a system of coupled loops and the measured frequencies
are not just functions of the properties of  the loop under observation. Instead,
the frequencies depend on the individual properties of all the loops and on the
collective properties of the system, such as the separation and spatial organization
of the loops \citep{Bogdan:1991, Luna:2008, Luna:2009, VanDoorsselaere:2008}.

\subsection{Coupling with Arcade Resonances}
\label{subsec:coupling_arcade}

MHD fast waves can propagate surprising distances within the corona.  Over the
typical lifetime of a coronal-loop oscillation, fast magnetic pressure waves can
propagate distances that are comparable to the radius of the Sun. For example,
\cite{Jain:2015} measure an oscillation lifetime of 5 minutes for several coronal
loops excited by a nearby flare. From the wave front launched from the flare site,
they estimate a coronal Alfv\'en speed of 2.5 Mm s$^{-1}$. Fast waves would therefore
travel 750 Mm during the loop oscillation's decay time. Since, the waves can travel
along and across field lines, the waves can efficiently propagate everywhere,
interacting with all of the flux structures within an active region. Furthermore,
the fast waves can travel back and forth many times across a typical active region;
thus, any resonances that are present should be excited. It should be no surprise
that the entire active region is observed to thrum and quiver after a flare, with
each flux system throbbing coherently.

{\it If the resulting modes of the arcade possess a frequency that is similar
to a frequency of the modes of a coronal loop that is embedded in the arcade, the
loop oscillations and the arcade oscillations will be coupled.} In our simple model,
such coincident frequencies only occur for the lateral overtones of the loop and
all such modes are of mixed character.  The fundamental kink mode maintains it's
purity as a loop oscillation; although we note that in the thin loop limit, the
wave's skin depth in the arcade becomes infinite because the wave's longitudinal
phase speed approaches the Alfv\'en speed of the arcade.  This behavior is a
consequence of the slab geometry \cite[e.g.,][]{Edwin:1982} and does not occur in
cylindrical geometry where the kink speed is intermediate between the Alfv\'en speeds
of the loop and the environment.

\subsection{Mode Misidentification}
\label{subsec:misidentification}

{\it Arcade resonances can masquerade as a loop resonance.} Once again, we emphasize
that without being able to observe the eigenfunction within the surrounding arcade
and along the loop at multiple positions, one can never be sure which modes have
been excited and which modes are being measured. For many events, such additional
information may not be forthcoming. However, in some cases, oscillations can be
observed within the surrounding arcade. For example, \cite{Allian:2019} used an
autocorrelation technique to identify the correlated oscillation of a background
of dim loops within the surrounding aracade. While they did not measure the axial
eigenfunction of arcade oscillations, they did demonstrate that the arcade as a
whole oscillated at nearly the same frequency as the loop. With further refinement,
it might be possible to use similar correlation techniques to identify the presence
or absence of nodes within the arcade by seeking for phase differences between
different parts of the arcade. Another possibility is to examine several loops
within the same arcade.  If the loops are well separated (further than a skin depth),
yet oscillate with the same phase, we may be observing a low-order arcade mode.
Similarly, if the two loops oscillate with anti-phase it is likely that the arcade
oscillation has a node in between the two loops.

A variety of previous efforts have detected multiple conincident frequencies of
oscillation for a single coronal loop \citep{Verwichte:2004, VanDoorsselaere:2007,
DeMoortel:2007, Li:2017, Duckenfield:2018}. Previously, such detections have been
implicitly assumed to be the consequence of the coexistence of modes corresponding
to different {\bf longitudinal} orders or overtones. Typically, two frequencies
have been detected, and the lowest frequency is assumed to be that of the fundamental
longitudinal mode ($m=1$) and the higher frequency concomitant with the first
longitudinal overtone ($m=2$). The ratio of frequencies of such longitudinal overtones
has been proposed as a diagnostic for variations along the loop in the mass density,
magnetic field strength, and loop radius \citep{Andries:2005, Goossens:2006, Dymova:2006,
Diaz:2007, McEwan:2008, Ruderman:2008, Verth:2008, Orza:2012, Jain:2012}. In observations,
it is rare for a majority of the length of the loop to be clearly visible, and hence
it is rare that one can verify which longitudinal modes have been excited by seeking
the presence of nodes. Thus, it is often impossible to know
if the two frequencies actually correspond to different longitudinal orders (different $m$).
Our work here suggests another possibility; the higher frequency oscillation could
correspond to an {\bf axial} overtone of the host arcade ($n>0$) instead of a longitudinal
overtone ($m>1$). Of course, both mechanisms could be, and probably are, at work
simultaneously. Since, Equation~\eqnref{eqn:freq_ratio} suggests that the effects of
axial overtones can be assessed through the aspect ratio of the arcade $W/L$, we strongly
urge that observations of an oscillation event attempt to characterize not only the 
length of the loop, but the width of the host arcade.

\subsection{Conclusions}
\label{subsec:conclusions}

Coronal seismology has long sought to model coronal loop oscillations as trapped
waves in a 1D waveguide that is isolated from other coronal structures. Typically
the selection of oscillation events is accomplished by looking for loops with high
brightness contrast that undergo large-amplitude motions that are easy to detect
and measure. Other than an obvious and justifiable bias towards loops that lack
overlapping and obscuring foreground and background features, the magnetic environment
of the loop is usually not a selection criterion. Here we have explored a few ways
in which the environment of the loop can break the basic assumptions implicit to
coronal seismology. In particular, we have shown that the lateral evanescent length, or
the skin depth, of kink oscillations can be remarkably long in the fluid surrounding
the loop. This means that one must be extremely judicious when selecting candidate
events for seismic analysis. The corona around the loop should be nearly featureless.

We have also explored how modes of the loop can couple to modes of the surrounding
arcade. When this happens, the waveguide takes on a multi-dimensional geometry. Our
model is 2D, but it is easy to imagine that the true waveguide could be fully 3D
\citep[e.g.,][]{Hindman:2015, Thackray:2017, Hindman:2018}. We also demonstrate that
observations made of the motion at a single position along the loop is insufficient
to allow unique identification of the mode of oscillation. Proper identification of
the mode is essential to correctly deduce the kink speed and other properties of the
loop. Once, again our results suggest that event selection needs to take the environment
of the loop into consideration. Any arcade resonances need to have sufficiently disparate
frequencies that coupling is weak.

Finally, in a broader context, coronal seismology needs more thorough testing. Currently,
we lack a basic understanding of the reliability and accuracy of seismic interferences.
The community needs to apply seismic techniques to a large sampling of loops for which
magnetic field strength and geometry estimates are available through spectroscopic and 
stereoscopic observations.


\acknowledgments

This work was supported by NASA through grants 80NSSC17K0008, 80NSSC18K1125,
80NSSC19K0267, and 80NSSC20K0193. R.J. would like to acknowledge the support of MSRC
(SoMaS), University of Sheffield (UK).




\end{document}